# HAMILTON-JACOBI THEORY AND THE EVOLUTION OPERATOR

J.F. Cariñena[1][*], X. Gràcia[2][†], E. Martínez[3][‡], G. Marmo[4][§],
M.C. Muñoz-Lecanda[2][¶], N. Román-Roy[2][‖]

[1] Departamento de Física Teórica, Facultad de Ciencias,
Universidad de Zaragoza, 50009 Zaragoza, Spain

[2] Departament de Matemàtica Aplicada 4, Universitat Politècnica de Catalunya.
Edifici C-3, Campus Nord UPC. C/ Jordi Girona 1. E-08034 Barcelona, Spain

[3] IUMA and Departamento de Matemática Aplicada, Facultad de Ciencias,
Universidad de Zaragoza, 50009 Zaragoza, Spain

[4] Dipartimento di Scienze Fisiche, Universitá Federico II di Napoli and INFN, Sez. Napoli.
Complesso Univ. di Monte Sant'Angelo, Via Cintia, 80126 Napoli, Italy

**Abstract.** We present a new setting of the geometric Hamilton-Jacobi theory by using the so-called *time-evolution operator* $\mathcal{K}$. This new approach unifies both the Lagrangian and the Hamiltonian formulation of the problem developed in [7], and can be applied to the case of singular Lagrangian dynamical systems.

*Keywords:* Hamilton–Jacobi equation, Lagrangian system, Hamiltonian system.
*PACS classification:* 02.40.Yy, 45.20.Jj.

## 1. Introduction

In a recent paper [7], a new geometric framework for the Hamilton–Jacobi theory has been presented. It is stated both for the Lagrangian and Hamiltonian formalisms of autonomous and non-autonomous regular dynamical systems, as well as for singular unconstrained mechanical systems. This framework has also been used to analyze the Hamilton-Jacobi problem in a different ambient such as nonholonomic mechanics [8, 25, 26], mechanics on Lie algebroids [28] and classical field theories [27].

The so-called *time-evolution $\mathcal{K}$-operator* (also known by some authors as the *relative Hamiltonian vector field* [32]) is a tool which has mainly been developed in order to study the Lagrangian and Hamiltonian formalisms for singular mechanical systems and their equivalence. It was first introduced in a non-intrinsic way in [3] as an "evolution operator" to connect both formalisms, as a refinement of the technique used in [15]. This operator was later defined geometrically in two different but equivalent ways [9, 17] for autonomous dynamical systems, and in [6] for the non-autonomous case.

[*]jfc@unizar.es
[†]xgracia@ma4.upc.edu
[‡]emf@unizar.es
[§]giuseppe.marmo@na.infn.it
[¶]mcml@ma4.upc.edu
[‖]nrr@ma4.upc.edu

The aim of this work is to show how the geometrical setting of the Hamilton-Jacobi equation can also be stated in terms of the evolution operator, so obtaining a formulation of the problem that comprises both the Lagrangian and the Hamiltonian formalism of the theory, and can be applied to the case of singular systems with constraints.

The organization of the paper is as follows: in section 2 we review the Lagrangian an Hamiltonian geometric Hamilton-Jacobi theory. The main features and properties of the evolution operator are presented in section 3. Finally, section 4 is devoted to present the main contribution of the paper; that is, the geometrical setting of the Hamilton-Jacobi problem using the evolution operator.

Manifolds and maps are assumed to be smooth. Sum over crossed repeated indices is understood. The natural projections of the tangent and the cotangent bundles will be denoted respectively by $\tau_Q : TQ \to Q$ and $\pi_Q : T^*Q \to Q$.

## 2. The Hamilton-Jacobi problem

Our formulation of the Hamilton-Jacobi problem is based on the idea of obtaining solutions of a second order differential equations by lifting solutions of an adequate first order differential equation. For details and proofs of the results stated in this section see [7].

### 2.1. Lagrangian Hamilton–Jacobi problem

A *Lagrangian dynamical system* is a pair $(TQ, L)$, where $L \in C^\infty(TQ)$ is the Lagrangian function of the system. Using the canonical structures in $TQ$ (i.e. the vertical endomorphism $S$, and the Liouville vector field $\Delta \in \mathfrak{X}(TQ)$) we construct the Lagrangian forms $\theta_L = dL \circ S \in \Omega^1(TQ)$ and $\omega_L = -d\theta_L \in \Omega^2(TQ)$, and the Lagrangian energy function $E_L = \Delta(L) - L \in C^\infty(TQ)$. Then, the Lagrangian dynamical equation is written

$$i(X_L)\omega_L = dE_L, \tag{1}$$

where the unknown is the vector field $X_L \in \mathfrak{X}(TQ)$. A solution of this equation is called a *Lagrangian vector field*. If the Lagrangian $L$ is regular (i.e. $\omega_L$ is a symplectic form) then there exists a unique solution, and its integral curves are holonomic ($X_L$ is a *Second Order Differential Equation*) and they satisfy the Euler-Lagrange equations.

Furthermore, we define the *Legendre transformation* associated with $L$ as the fiber derivative of the Lagrangian function $\mathcal{F}L \colon TQ \to T^*Q$, i.e $\langle \mathcal{F}L(v), w \rangle = \frac{d}{ds}L(v+tw)|_{s=0}$, for $v, w \in TQ$, with $\tau_Q(v) = \tau_Q(w)$.

For a regular Lagrangian $L \in C^\infty(TQ)$, the *generalized Lagrangian Hamilton–Jacobi problem* consists in finding a vector field $X \colon Q \to TQ$ such that, if $\gamma \colon \mathbb{R} \to Q$ is an integral curve of $X$, then $X \circ \gamma = \dot\gamma \colon \mathbb{R} \to TQ$ is an integral curve of $X_L$; that is,

$$X \circ \gamma = \dot\gamma \implies X_L \circ \dot\gamma = \overline{X \circ \gamma}^{\,\cdot}.$$

$X$ is said to be a *solution to the generalized Lagrangian Hamilton–Jacobi problem*.

**Theorem 1.** *Let L be a regular Lagrangian and $X \in \mathfrak{X}(Q)$. The following assertions are equivalent:*

1. *X is a solution to the generalized Lagrangian Hamilton–Jacobi problem.*
2. *X and $X_L$ are X-related; that is, $X_L \circ X = TX \circ X$.*
3. *The submanifold $\operatorname{Im} X \subset TQ$ is invariant by the Lagrangian vector field $X_L$ (that is, $X_L$ is tangent to the submanifold $X(Q)$).*
4. *X satisfies the equation $i(X)(X^*\omega_L) = d(X^*E_L)$.*

The standard version of the Hamilton–Jacobi problem is by obtained considering a simplified problem. Given a Lagrangian function $L \in C^\infty(TQ)$, the *Lagrangian Hamilton–Jacobi problem* consists in finding solutions $X$ to the generalized Lagrangian Hamilton–Jacobi problem satisfying that $X^*\omega_L = 0$.

**Theorem 2.** *Let $X \in \mathfrak{X}(Q)$ such that $X^*\omega_L = 0$. The following assertions are equivalent:*

1. *X is a solution to the Lagrangian Hamilton–Jacobi problem.*
2. *$d(X^*E_L) = 0$.*
3. *$\operatorname{Im} X$ is a Lagrangian submanifold of $TQ$ invariant by $X_L$.*
4. *The integral curves of $X_L$ with initial conditions in $\operatorname{Im} X$ project onto the integral curves of $X$.*

As $0 = X^*\omega_L = -X^*d\theta_L = -d(X^*\theta_L)$, then there exist an open neighborhood $U \subset Q$ of every point of $Q$, and a function $W$ such that $X^*\theta_L = dW$ (in $U$). The coordinate expression of the second characterization, $E(q^i, \partial W/\partial q^i) = $ constant, gives the classical form of the Hamilton–Jacobi equation in the Lagrangian formalism [31].

## 2.2. Hamiltonian Hamilton–Jacobi problem

A *Hamiltonian system* is a pair $(T^*Q, H)$, where $H \in C^\infty(T^*Q)$ is the *Hamiltonian function* of the system. Using the canonical symplectic form $\omega = -d\theta \in \Omega^2(T^*Q)$ on the cotangent bundle $T^*Q$, the *Hamiltonian dynamical equation* reads

$$i(Z_H)\omega = dH . \tag{2}$$

where the unknown is the vector field $Z_H \in \mathfrak{X}(T^*Q)$, which is called the *Hamiltonian vector field*. Its integral curves are the trajectories of the system, and they satisfy the *Hamilton equations*.

Given a Hamiltonian vector field $Z_H \in \mathfrak{X}(T^*Q)$, the *generalized Hamiltonian Hamilton–Jacobi problem* consists in finding a vector field $X \colon Q \to TQ$ and a 1-form $\alpha \colon Q \to T^*Q$ such that, if $\gamma \colon \mathbb{R} \to Q$ is an integral curve of $X$, then $\alpha \circ \gamma \colon \mathbb{R} \to T^*Q$ is an integral curve of $Z_H$; that is,

$$X \circ \gamma = \dot{\gamma} \implies \overline{\dot{\alpha \circ \gamma}} = Z_H \circ (\alpha \circ \gamma) . \tag{3}$$

This condition is equivalent to

$$Z_H \circ \alpha = T\alpha \circ X .$$

Furthermore, if $(X, \alpha)$ satisfies the condition (3), then we obtain the following relation between these elements

$$X = T\pi_Q \circ Z_H \circ \alpha = \mathcal{F}H \circ \alpha,$$

where $\mathcal{F}H$ is the fibre derivative of $H$.

A *solution* to the generalized Hamiltonian Hamilton–Jacobi problem for $Z_H$ is a 1-form $\alpha \in \Omega^1(Q)$ such that, if $\gamma: \mathbb{R} \to Q$ is an integral curve of $X = T\pi_Q \circ Z_H \circ \alpha$, then $\alpha \circ \gamma: \mathbb{R} \to T^*Q$ is an integral curve of $Z_H$; that is,

$$T\pi_Q \circ Z_H \circ \alpha \circ \gamma = \dot{\gamma} \implies \dot{\overline{\alpha \circ \gamma}} = Z_H \circ (\alpha \circ \gamma).$$

**Theorem 3.** *Let $\alpha \in \Omega^1(Q)$. The following assertions are equivalent:*

1. *$\alpha$ is a solution to the generalized Hamiltonian Hamilton–Jacobi problem.*
2. *The submanifold $\operatorname{Im}\alpha \subset T^*Q$ is invariant under the flow of the vector field $Z_H$ (that is, $Z_H$ is tangent to the submanifold $\operatorname{Im}\alpha$).*
3. *$\alpha$ satisfies the equation $i(X)d\alpha = -d(\alpha^* H)$, where $X = T\pi_Q \circ Z_H \circ \alpha$ is the vector field associated with $\alpha$.*

As in the Lagrangian formalism, we consider a simpler case which leads to the standard version of the Hamilton-Jacobi problem. Thus, given a vector field $Z_H \in \mathfrak{X}(T^*Q)$, the *Hamiltonian Hamilton–Jacobi problem* consists in finding a *closed* 1-form $\alpha \in \Omega^1(Q)$ which is a solution to the generalized Hamiltonian Hamilton–Jacobi problem. Notice that $d\alpha = 0$ is equivalent to $\alpha^*\omega = 0$.

**Theorem 4.** *Let $\alpha$ be a closed 1-form in $Q$. The following assertions are equivalent:*

1. *$\alpha$ is a solution to the Hamiltonian Hamilton–Jacobi problem.*
2. *$d(\alpha^* H) = 0$ (that is, $\alpha^* H$ is locally constant).*
3. *$\operatorname{Im}\alpha$ is a Lagrangian submanifold of $T^*Q$ invariant by $Z_H$.*
4. *The integral curves of $Z_H$ with initial conditions in $\operatorname{Im}\alpha$ project onto the integral curves of $X = T\pi_Q \circ Z_H \circ \alpha$.*

Since $\alpha$ is closed, there is a function $W$ defined in an open neighborhood $U \subset Q$ of every point and such that $\alpha = dW$ (in $U$). The coordinate expression of the second characterization $H(q^i, \partial W/\partial q^i) = $ constant, gives the classical form of the Hamilton–Jacobi equation in the Hamiltonian formalism.

## 3. The evolution operator

The above results allow to study Hamilton-Jacobi theory for regular Lagrangians and the corresponding Hamiltonian problem, and provides an understanding of the situation whenever we have alternative Lagrangian descriptions for the same dynamical vector field [7]. However, it is well known that many interesting physical systems are defined by gauge invariant Lagrangians, a fact that forces the Lagrangian to be singular. It is therefore interesting to look for a more general formalisms where one can study easily the dynamics defined by singular Lagrangians and the Hamilton-Jacobi theory. One of such formalisms is provided by the so-called time-evolution operator $\mathcal{K}$.

The definition given here for the evolution operator is based on the concept of *section along a map* [5, 11, 12, 30], of which this operator is a particular case. We recall that a vector field $Z$ along a map $F: N \to M$, is a map $Z: N \to TM$ such that $\tau_M \circ Z = F$. The set of vector fields along $F$ will be denoted $\mathfrak{X}(F)$. An integral curve of $Z \in \mathfrak{X}(F)$ is a curve $\gamma : \mathbb{R} \to N$ such that $Z \circ \gamma = \frac{d}{dt}(F \circ \gamma)$.

To every vector field $Z \in \mathfrak{X}(F)$ we can associate a $F^*$-derivation $i(Z)$ of degree $-1$ determined by
$$(i(Z)\omega)_n(v_1, \ldots, v_p) = \omega_{F(n)}(Z(n), T_n F(v_1), \ldots, T_n F(v_p))$$
for every $p$-form $\omega$ in $M$, every $n \in N$ and $v_1, \ldots, v_p \in T_n N$. In this way an equation between vector fields $Y, Z$ along some maps of the form $TG \circ Z = Y \circ H$, where $G$ and $H$ are maps, is equivalent to the relation $i(Z) \circ G^* = H^* \circ i(Y)$. See [29] for more details about $F^*$-derivations.

The *time-evolution operator* $\mathcal{K}$ associated with a Lagrangian $L : TQ \to \mathbb{R}$ is a vector field along the Legendre transformation, i.e. $\mathcal{K} : TQ \to T(T^*Q)$ with $\tau_{T^*Q} \circ \mathcal{K} = \mathcal{F}L$, satisfying the following two conditions:

1. (Dynamical condition):
$$i(\mathcal{K})\omega = dE_L . \tag{4}$$

2. (Second-order condition):
$$T\pi_Q \circ \mathcal{K} = \mathrm{Id}_{TQ}, \tag{5}$$
where $\pi_Q : T^*Q \to Q$ is the canonical projection.

The existence and uniqueness of this operator is studied in [17], and it can be shown that it is explicitly given by $\mathcal{K} = \chi \circ dL$, where $\chi : T^*TQ \to TT^*Q$ is the canonical isomorphism [35]. Its local expression is
$$\mathcal{K}(q, v) = v^i \left.\frac{\partial}{\partial q^i}\right|_{\mathcal{F}L(q,v)} + \frac{\partial L}{\partial q^i} \left.\frac{\partial}{\partial p_i}\right|_{\mathcal{F}L(q,v)} . \tag{6}$$

By definition $\varphi : \mathbb{R} \to TQ$ is an integral curve of $\mathcal{K}$ if $T(\mathcal{F}L) \circ \dot{\varphi} = \mathcal{K} \circ \varphi$, where $\dot{\varphi} : \mathbb{R} \to T(TQ)$ is the prolongation of $\varphi$ to the tangent bundle $T(TQ)$ of $TQ$. Moreover, as a consequence of the second-order condition (in the item 3), we have that $\varphi = \dot{\phi}$, for $\phi : \mathbb{R} \to Q$, that is, $\varphi$ is holonomic. We have the following commutative diagram

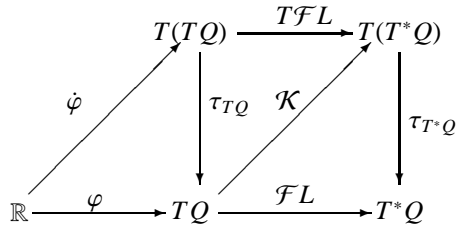

In coordinates the integral curves of $\mathcal{K}$ are characterized by the equations
$$\dot{x} = v , \qquad \dot{p} = \frac{\partial L}{\partial x} , \qquad \text{and} \qquad p = \frac{\partial L}{\partial v}, \tag{7}$$
which are clearly equivalent to the Euler-Lagrange equations for $L$

The most relevant properties of this operator are the following [3, 9, 17, 18]:

- If there exists a vector field $X_L$ on $TQ$ which is a solution to the equation (1), then $\varphi : \mathbb{R} \to TQ$ is an integral curve of $X_L$ if, and only if, it is an integral curve of $\mathcal{K}$. As a direct consequence of this fact, the relation between $\mathcal{K}$ and $X_L$ is

$$T(\mathcal{F}L) \circ X_L = \mathcal{K} \,. \tag{8}$$

- If there exists a Hamilton vector field $X_H$ on $\mathcal{F}L(TQ) \subseteq T^*Q$ which is a solution to the Hamilton equations in the Hamiltonian formalism associated to the Lagrangian system $(TQ, L)$, then $\psi : \mathbb{R} \to T^*Q$ is an integral curve of $X_H$ if, and only if,

$$\dot\psi = \mathcal{K} \circ T\pi_Q \circ \dot\psi \,. \tag{9}$$

As a consequence, the relation between $\mathcal{K}$ and $X_H$ is

$$X_H \circ \mathcal{F}L = \mathcal{K} \,. \tag{10}$$

Equations (8), (9) and (10) show how the Lagrangian and Hamiltonian descriptions can be unified by means of the operator $\mathcal{K}$.

Other relevant results obtained using this operator are:

- The equivalence between the Lagrangian and Hamiltonian formalisms is proved by means of this operator in the following way: there is a bijection between the sets of solutions of Euler-Lagrange equations and Hamilton equations.

- The complete classification of constraints appearing in the constraint algorithms for singular dynamical systems is achieved. All the Lagrangian constraints can be obtained from the Hamiltonian ones using the $\mathcal{K}$-operator in the following way: if $\xi \in C^\infty(T^*Q)$ is a Hamiltonian constraint, then $i(\mathcal{K})d\xi$ is a Lagrangian constraint.

Most of these results have also been generalized for higher-order Lagrangian systems [4, 10, 12, 23, 24], and for the case of more general types of singular differential equations on manifolds [18]. The evolution operator $\mathcal{K}$ has also been applied for analyzing gauge symmetries, Noether's theorem and other structures for singular systems [14, 16, 19, 20, 21, 22], and to study Lagrangian systems whose Legendre map degenerates on a hypersurface [32, 33]. Furthermore, although a covariant description of this operator was not available, it has also been used to study several characteristics of some physical models in field theory, namely the *bosonic string* [1, 2, 19]. Later on, the definition and some properties of the operator $\mathcal{K}$ have been carried out for field theories [13, 34].

## 4. The Hamilton-Jacobi problem and the evolution operator

Let $L$ be a Lagrangian on $TQ$, and let $\mathcal{K}$ be the associated time-evolution operator. Our aim is to characterize the solutions of the generalized Hamilton-Jacobi problem in terms of $\mathcal{K}$. In this way, we will be able to find solutions to such problem even for a singular Lagrangian $L$.

For the following result, we recall that a vector field $Z$ along a map $F : N \to M$ is tangent to a submanifold $P \subset N$ if, for every $p \in P$, we have that $Z(p) \in T_pF(T_pP)$. In other words, the restriction $Z|P$ takes values tangent to $F(P)$.

**Theorem 5.** *Let $X$ be a vector field on the manifold $Q$ and consider the 1-form $\alpha = \mathcal{F}L \circ X$. The following conditions are equivalent:*

1. *If $\gamma$ is an integral curve of $X$ then $\dot{\gamma}$ is an integral curve of $\mathcal{K}$.*
2. *$\mathcal{K}$ is tangent to $\mathrm{Im}(X)$.*
3. *$T\alpha \circ X = \mathcal{K} \circ X$.*
4. *$i(X)d\alpha + d(X^*E_L) = 0$.*
5. *$i(X)(X^*\omega_L) - d(X^*E_L) = 0$.*

**Proof:**

[$1 \Rightarrow 2$] Let $v \in TQ$ be a point in $\mathrm{Im}(X)$, and set $q = \tau_Q(v)$, so that $v = X(q)$. Let $\gamma$ be the integral curve of $X$ starting at $q$, $\gamma(0) = q$. It follows that the initial velocity of this curve is $\dot{\gamma}(0) = X(\gamma(0)) = X(q) = v$. The curve $\dot{\gamma}$ is an integral curve of $\mathcal{K}$, from where

$$\mathcal{K}(v) = (\mathcal{K} \circ \dot{\gamma})(0) = \frac{d}{dt}(\mathcal{F}L \circ \dot{\gamma})(0) = \frac{d}{dt}(\mathcal{F}L \circ X \circ \gamma)(0),$$

that is, the vector $\mathcal{K}(v)$ is the tangent at $t = 0$ to the curve $\mathcal{F}L \circ X \circ \gamma$, which is a curve in $\mathcal{F}L(\mathrm{Im}(X))$.

[$2 \Rightarrow 3$] If $\mathcal{K}$ is tangent to $\mathrm{Im}(X)$ then $\mathcal{K} \circ X$ takes values tangent to $\mathrm{Im}(\mathcal{F}L \circ X) = \mathrm{Im}(\alpha)$. Thus $\mathcal{K} \circ X = T\alpha \circ Y$, for some vector field $Y$ on $Q$. But taking the projection, we have on one hand $T\pi_Q \circ \mathcal{K} \circ X = \mathrm{id}_{TQ} \circ X = X$, and on the other

$$T\pi_q \circ T\alpha \circ Y = T(\pi_Q \circ \alpha) \circ Y = T\mathrm{id}_Q \circ Y = \mathrm{id}_{TQ} \circ Y = Y.$$

Thus $Y = X$ and the result follows.

[$3 \Rightarrow 1$] Let $\gamma$ be an integral curve of $X$, that is $X \circ \gamma = \dot{\gamma}$. Composing the equation $T\alpha \circ X = \mathcal{K} \circ X$ with $\gamma$ we get $T\alpha \circ \dot{\gamma} = \mathcal{K} \circ \dot{\gamma}$, and taking into account the definition of $\alpha$ we have

$$\mathcal{K} \circ \dot{\gamma} = T\alpha \circ \dot{\gamma} = \frac{d}{dt}(\alpha \circ \gamma) = \frac{d}{dt}(\mathcal{F}L \circ X \circ \gamma) = \frac{d}{dt}(\mathcal{F}L \circ \dot{\gamma}),$$

so that $\dot{\gamma}$ is an integral curve of $\mathcal{K}$.

[$3 \Rightarrow 4$] Taking the pullback by $X$ of the equation $i(\mathcal{K})\omega = dE_L$ we get $X^*i(\mathcal{K})\omega = d(X^*E_L)$. Since the relation $T\alpha \circ X = \mathcal{K} \circ X$ is equivalent to $i(X) \circ \alpha^* = X^* \circ i(\mathcal{K})$, we have that $i(X)(\alpha^*\omega) = d(X^*E_L)$. Finally, taking into account that $\alpha^*\omega = -d(\alpha^*\theta) = -d\alpha$, we get $-i(X)d\alpha = d(X^*E_L)$.

[$4 \Rightarrow 3$] Let $X$ be a vector field satisfying $i(X)d\alpha + d(X^*E_L) = 0$. Taking the pullback of the equation $i(\mathcal{K})\omega = dE_L$ by $X$ we have $X^*i(\mathcal{K})\omega - d(X^*E_L) = 0$. From this two equations we get $X^*i(\mathcal{K})\omega - i(X)\alpha^*\omega = 0$, or equivalently $i(\mathcal{K} \circ X - T\alpha \circ X)\omega = 0$.

We next show that this condition implies $\mathcal{K} \circ X - T\alpha \circ X = 0$. Denote by $D \in \mathfrak{X}(\alpha)$ the vector field along $\alpha$ given by $D = T\alpha \circ X - \mathcal{K} \circ X$, so that $i(D)\omega = 0$. This equation is explicitly written as $\omega(D(q), T_q\alpha(w)) = 0$, for every $q \in Q$ and $w \in T_qQ$. The vector field $D$ along $\alpha$ is vertical

$$T\pi_Q \circ D = T\pi_Q \circ T\alpha \circ X - T\pi_Q \circ \mathcal{K} \circ X = X - X = 0,$$

and since the vertical subbundle is Lagrangian for $\omega$, it follows that $\omega(D(q), V) = 0$ for every vertical vector $V \in T_{\alpha(q)}T^*Q$. Finally, every vector $W \in T_{\alpha(q)}T^*Q$ can be written $W = T\alpha(w) + V$, with $w = T\pi_Q(W)$ and $V$ vertical, from where

$$\omega(D(q), W) = \omega(D(q), T\alpha(w)) + \omega(D(q), V) = 0,$$

and the result follows.

[4 $\Leftrightarrow$ 5] This equivalence follows from $X^*\omega_L = X^*\mathcal{F}L^*\omega = \alpha^*\omega = -d\alpha$. $\square$

In general, the time-evolution operator $\mathcal{K}$ will not be tangent to the full space $TQ$, and the system will have constraints. Applying the constraint algorithm to our system, we can find a submanifold $M \subset TQ$ such that $\mathcal{K}$ is tangent to it, which is known as the final constraint submanifold. In such a case we have to restrict our vector fields, and we must consider vector fields $X$ taking values on the final constraint manifold.

The standard Hamilton-Jacobi problem is obtained by considering a simplification of our general problem. The equations in the item 5 are nonlinear in $X$ and are difficult to solve, thus we may simplify the problem by looking for a particular kind of solutions, namely those vector fields $X$ such that associated 1-form $\alpha$ is closed, that is, $d\alpha = 0$. From here it follows that the energy is constant on the image of the section $X$, that is, $X^*E_L$ = constant. This equations are said to be the *Hamilton-Jacobi equation in the mixed formalism*

$$d\alpha = 0 \quad \text{and} \quad X^*E_L = \text{constant}.$$

When we realize that $\alpha = X^*\theta_L$ then we get the equations

$$d(X^*\theta_L) = 0 \quad \text{and} \quad X^*E_L = \text{constant},$$

which is the Hamilton-Jacobi equation in the Lagrangian formalism. Furthermore, assuming that the energy is $\mathcal{F}L$-projectable, so that a Hamiltonian function $H$ exists such that $\mathcal{F}L^*H = E_L$, then from this relation we get the equations

$$d\alpha = 0 \quad \text{and} \quad \alpha^*H = \text{constant},$$

which is the Hamilton-Jacobi equation in the Hamiltonian formalism.

## Acknowledgements

We acknowledge the financial support of the *Ministerio de Educación y Ciencia*, projects MTM2005-04947, MEC-DGI MTM2006-10531, MTM2008-00689/MTM and MTM2008-03606-E/MTM, FPA-2003-02948 and CO2-399.